\documentclass[runningheads,12pt]{llncs}
\usepackage{setspace}
\usepackage[left=2.5cm, right=2.5cm, top=2.5cm]{geometry}
\usepackage{graphicx}
\usepackage{amsmath}
\usepackage{amssymb}
\usepackage{mathrsfs}
\usepackage{booktabs} 
\usepackage{subcaption}
\usepackage{lineno}

\usepackage{hyperref}
\usepackage{graphicx}
\usepackage{listings}
\usepackage{epstopdf}
\usepackage{color}
 
\usepackage[utf8]{inputenc}
\usepackage{amsmath}
\usepackage{algorithm}
\usepackage{algpseudocode}

\begin{document}
\title{Unraveling Managerial Tangents in Firm Disclosure: Concealing Issues or Being Exposed?}
\titlerunning{  }
%
\author{Xuan Zhou\inst{1} \and
Yushen Huang\inst{2} }
\authorrunning{X. Zhou et al.}
%
\institute{University of Massachusetts Amherst, Amherst, MA, 01375, USA \email{\{xuan.zhou\}@umass.edu} \and Stony Brook University, Stony Brook NY 11790, USA \\ \email{\{yushen.huang\}@stonybrook.edu} \ }
\maketitle              
\begin{abstract}

Earnings calls influence stock prices and are traditionally analyzed using sentiment and linguistic traces. Our research introduces a "Topic-Switching Index," a novel metric quantified through the transformer model FinBERT, to measure managerial evasion during Q$\&$A sessions in earnings calls. We find a negative correlation between this index and subsequent stock prices, indicating that investors penalize managerial evasiveness. This study is the first to quantify such evasive tactics, adding a new dimension to how earnings calls are understood and suggesting that topic shifting is an overlooked but significant factor. We also show the predictability of the index under three different classifier models and it stands out in all circumstances.
\end{abstract}
\pagebreak
\section{Introduction}
An earnings call is a conference call between the management team of a public company and analysts to discuss the company’s financial conditions after a given reporting period, usually after a quarter or a year. 
In each earnings call, the representee of the management team, most of the time, the CEO, will firstly show the current financial achievements of the company and then describe the plan for the upcoming quarter. 
Next, each analyst attended can ask one question with a follow up question, which are ought to be answered by the manager no matter how awkward they could be. 
Therefore, the earnings call is an important event as the analysts can ask critical questions to the manager and the manager's response can reveal more information that is not fully covered in the financial reports. In fact, ending up no answers to such questions is also a way of reply.

\begin{figure}
    \centering
    \includegraphics[width=\linewidth]{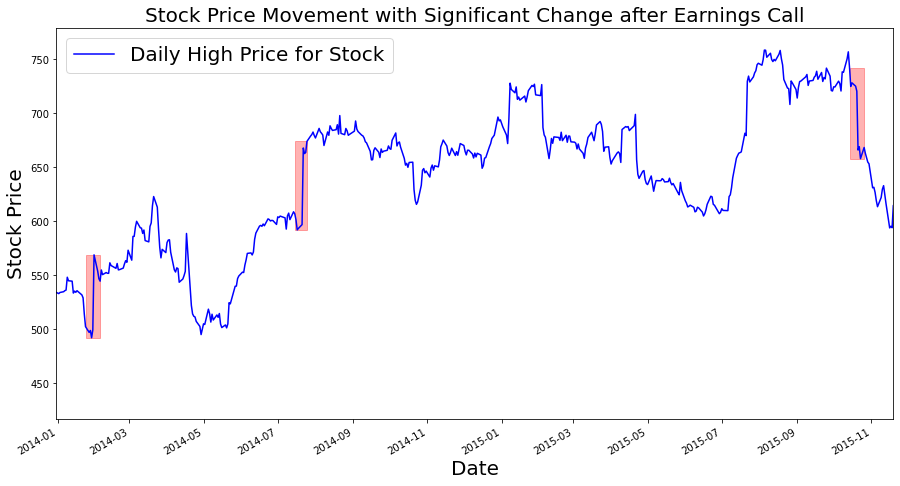}
    \caption{Stock Price movement of Chipotle Mexican Grill}
    \label{fig:Stock_Price_Sample}
\end{figure}

From the conference, the investors can gather new information and adjust their investment decisions accordingly. As shown in Figure \ref{fig:Stock_Price_Sample}, we can see the stock price of Chipotle Mexican Grill has three big jumps on three earnings call dates relatively.
It is indicative for us to conduct research on how does the earnings call affect the movement of campany's stock price and looking into the earnings call transcripts for clues.
Moreover, \cite{matsumoto2011makes} demonstrate that discussion periods are relatively more informative than presentation periods.
So our reaserach is mainly focused on analyzing the Q$\&$A session of the earnings call transcripts.
Those transcripts are provided for free by the host companies as well as on some third-party websites. In our paper, we collected those text from a website called seeking alpha.

However, there are mainly two concerns for using earnings call transcript to forecast stock price movement. 1. What factors should we consider in the earnings call to forecast the tendency? 2. How can we quantify those factors?

In literature, transcripts affect the stock price mainly in two ways, either from the sentiment aspect or the linguistic traces.
So far, plenty of literature has proved that the sentiment contained have significant impact on stock price movement \cite{chen2018manager,kimbrough2005effect,garcia2023colour,ke2019predicting} 
Additionally, the sentiment from managers and analysts does not share the same weight in investors' decision-making process. In \cite{chen2018manager}, it is found that intraday prices react significantly to analyst tone, but not to management tone, for the full duration of the discussion. This effect strengthens when the analyst's tone is relatively negative. 
To the investors, when managers show unwillingness to answer questions directly, it also signals potential bad performance of the company~\cite{hollander2010does}. 
The linguistic complexity in manager's speech itself also indicates that the manager is trying to evade from the question, which will be reflected in the stock price later ~\cite{youngs2020linguistic,chi2022does,bushee2021analysts,li2008annual,bloomfield2008discussion,guay2016guiding,dyer2017evolution} talking about the influence of the linguistic complexity in finance. 
Nevertheless, it is then pointed out in \cite{bushee2018linguistic} that linguistic complexity commingles two latent components—obfuscation and information—that are related to information asymmetry in opposite directions. By substract the complexity of analysts scripts from that of managers', this paper comes up with a novel approach to offset the language complexity required to understand the certain industry, for example, the terminology used. 
\cite{gow2021non} quantify a factor of whether the manager answers the analysts' question. By detecting the presence of key phrases in the response, which is defined to have three certain forms, they classify a managerial response to a question as a non-answer using regular expression. It is also shown in the paper that 11 percent of the managers do not answer the question from the analyst.
~\cite{lee2016can} show that a lack of spontaneity is negatively associated with the market reaction to the call, by using a measure of the adherence of transcripts to prepared scripts.

In conclusion, most of the previous studies employing natural language processing techniques on earnings call transcripts have focused on the emotional reactions or the linguistic way of responding, which both are proven to be reliable indicators of market expectations. 
However, studies based on sentiment analysis have a common tendency to overlook the potential significance of managers' responses. 
Managers' scripts, althogh much longer than analysts', are proved to be less useful in such researches. \cite{davis2015effect} find that tone is significantly associated with manager-specific factors such as early career experiences, and involvement in charitable organizations. Managers can make themselves sound as optimistic while the fact does not appear to be.
While being more useful in sentiment analysis, the scripts from analysts are usually very short and brief as they are only allowed to have one question and a follow up question at most, let alone make comments on the managers' answers. So the sentiment detected from analysts' transcripts onle can be less accurate and hard to tell difference quantitatively.
On the other hand, researches focused on linguistic analysis of managers' answers can be also biased because of managers’ own wording behavior as they can be trained to use less complex language.

In contrast, our research highlights the importance of revealing more information hidden under managers' answers. 
While managers can manipulate the sentiment expressed through training and preparation in advance of a conference, and adding contents that sound positive in the speech like good aspects of the company's performance, topic shifting remains a challenging behavior to disguise. 
When managers encounter questions for which they cannot provide positive or satisfying answers, their best option is often to evade the question altogether. 
Our study quantifies this topic shifting behavior and examines the market's response to it.

Our main assumption is that from the way the managers answering questions, the investors have their own thoughts on whether the company is really as what it is as described during the earnings call, or the managers deliberately hide some potential issues. Those suspicions will be later reflected in the stock market. 
We assume that during an earnings call, if the topics contained in the managers' responses in the Q$\&$A session do not align or significantly deviate from the themes present in analysts' questions, it indicates a deliberate topic diversion tactic. 
Such diversions can be interpreted as strategies adopted by managers to circumvent questions that may be difficult to address or those that might not yield affirmative reactions from analysts and investors. 
Furthermore, this topic shifting allows managers to portray a more positive and upbeat sentiment in their transcripts. 
The effect of sentiment is widely researched, and some third-party websites provide sentiment score of earnings call transcript for investors who have membership, so we have good reason to hypothesize that the managers will try to at least sound as positive as possible. 
While they need to face the questions and sound confident at the same time, it is common that the manager evades from tough question like spending more time adding details that are not necessarily related to the question itself. 
Conversely, when the topics of questions and answers align well, we believe that the manager is confident in responding in a positive and constructive manner, suggesting the company is in a good shape and do not have suspicious issues or the current problems faced by the company are well under control. 

We further hypothesize that investors will detect the behavior of managers switching topics to dodge questions, and this will be reflected in stock price movement later on. 
More specifically, the act of topic shifting during an earnings call Q$\&$A session will have a negative impact on the stock price of the host company, while candid responses can have a positive impact.

In our research, which is the first in this area to put forth this hypothesis, we have designed a Topic Shifting Index. 
We use a transformer method called Bidirectional Encoder Representation Transformer (BERT) ~\cite{devlin2018bert} to quantify this topic switching factor which perform better than traditional methods mentioned in lierature so far in sentiment analysis area \cite{liu2021finbert,araci2019finbert}. 
This metric quantifies the degree of alignment between analysts' questions and managers' responses. According to our speculation, it has a negative correlation with stock prices post earnings call, which is proved to be true in this study and our model based on this feature outperforms the latest model in the literature so far. Our method allows us to distinguish from previous literature on mamager's attempt to evade from questions in firm disclosure in that previous literature only identify cases that managers do not answer questions at all and show liguistic traces while ours includes circumstances that managers do not fully answer questions or in a indirect way. 

Our research contribute to literature in methodology also in that we manage to separate every analyst's questions from the others'. 
Unlike previous literature, which quantify the features interested based on all the analysts or/and all the managers in one earnings call as a whole, we pair each analyst with the answers they get in each call and get the score. Then we calculate the average score for that conference. 
It can help us get a more accurate score and is also more clear and straightforward when checking the reliability of our topic switching index under different context. 
There are circumstances when the first analyst had an easy question and the manager replied with confidence, but then the second analysts raised a tough one and the manager answered using the same material covered in the first answer again. 
If we do not separate the analysts from each other, it is hard to get a credible score as the manager's answers do match with the topics in the question pool. 
With pairing every analyst with the answers they get whether they have a follow-up question or not, we assign them with equal weight during one earnings call and will avoid the bias from puting too much weight on questions with long answers.

Our findings challenge the argument made by \cite{chen2018manager}
that analysts are the participants on earnings calls whose comments move stock prices during the discussion, showing that managers and their answering is also valuable indicator and detected by the market.
Our study uncovers an additional layer of information embedded in managers' responses, shedding light on their communication strategies and potential attempts to evade certain questions that they are supposed to answer directly.
At the same time, we do not overlook the role played by the analysts in the earnigs call but include the information in their transcripts as well. By comparing the similarity of questions and answers in pair under the whole context, we extract as much information as possible from both parties.
Moreover, our research makes a significant contribution to this area by quantifying topic shifting behavior and investigating its impact on stock prices. 
While prior studies have explored linguistic cues and sentiment in earnings calls, our study is the first to extract topic shifting feature and prove that the managers do try to avoid answering tough questions with adding less related contents in the answers to make themselves sound good and redirect the topics. 
By evaluating the relationship between detected instances of topic shifting and subsequent stock price movements, we also provide empirical evidence on how market participants perceive and respond to this hidden information released by managers.
This contribution expands the understanding of the information dynamics within earnings calls transcripts and offers valuable insights to analysts and institutional investors, enabling them to have better forecast and form financial strategies accordingly from these fiscal communication events.

The remainder of this paper is structured as follows: 
Section 2 presents our methodology for detecting topic shifting by managers and outlines the measures employed to assess market reactions. 
Section 3 provides an overview of the data collection and preprocessing methods, including the selection of earnings call transcripts and the application of NLP techniques. 
Section 4 discusses the empirical results and their implications. 
Finally, Section 5 summarizes the findings, discusses limitations, and suggests avenues for future research.

\section{Methdology}
In this section, we explain our methodology to calculate the Topic Switching Index and how to use it to predict the tendency of the stock price. 
\subsection{Notation}
We start by introducing the notation of the stock price and its relative definition. Those definitions and notations follow from \cite{medya2022exploratory}.  \\ \\
\ \
    Let $\mathbb{C} = \{ C_1, C_2,\cdots, C_m \} $ be the set of $m$ companies and $S_{d}^{c}$ be the stock high price for company $c$ at day $d$. We also denote the earnings call transcripts of the company throughout all the periods as the set $\mathbb{T}^{c} = \{ T_{d_1}^{c}, T_{d_2}^{c},\cdots, T_{d_t}^{c} \}$. Next, let us define the tendency of the stock price movement by checking that if the earnings call happens on day $d$, whether the stock price from day $d+1$ is greater or smaller than the stock price on day $d - 1$.
    \begin{definition} \label{stock_price_tendency}
    \textbf{Value Based Label Function(VBL)}. We define the label function $y(T_d^c) \in \{ -1, 1 \}$ for a transcript T$_d$ of a company  on the day $d$ as follows:
    \begin{align*}
        y(T_d^c) = \left \{ \begin{array}{cc}
         1    & \mbox{ If } T_{d+1}^c \geq T_{d-1}^c \\
        -1   &  \mbox{ If } T_{d+1}^c < T_{d-1}^c
        \end{array} \right . 
    \end{align*}
    \end{definition}
    \ \
 As mentioned above, \textbf{Definition} \ref{stock_price_tendency} corresponds to whether the stock price on the day post the call increases compared with one day before; however, in reality, whether the investment makes a profit depends on the current risk-free rate. Hence the tendency is positive only if the stock price increase rate is larger or equal to a threshold $\tau$. Next let us define the relative stock price movement as the following:
\begin{definition} \label{stock_price_tendency_relative}
    \textbf{Value Based Label Function(VBL)}. We define the label function $y_r(T_d^c) \in \{ -1, 1 \}$ for a transcript T$_d$ of a company  on the day $d$ as follows, where $\tau$ is the risk-free rate on day $d$:
    \begin{align*}
        y_r(T_d^c) = \left \{ \begin{array}{cc}
         1    & \mbox{ If } \frac{T_{d+1}^c - T_{d-1}^c}{T_{d-1}^c}  \geq \tau \\
        -1   &  \mbox{ If }  \frac{T_{d+1}^c - T_{d-1}^c}{T_{d-1}^c}  < \tau 
        \end{array} \right . 
    \end{align*}
\end{definition}
\subsection{BERT and Transformer Based Model}
For both the Topic-Switching Feature and the Benchmark Feature, they are based on BERT~\cite{devlin2018bert} which is a Transformer Model~\cite{vaswani2017attention}. 
\subsubsection{Transformer Model}:
The transformer model is based on self-attention which relates to the different positions of a sequence and converts it to a vector that includes that information. More specifically, the Transformer Model has the following stage: encoder and decoder~\cite{cho2014learning,bahdanau2014neural,sutskever2014sequence} stage, and then it will go over the linear and softmax function which gives the final output. We will dig into the details of the model after familiarizing with some definitions:
\begin{definition}
\textbf{Scaled Dot-Product Attention}: Given an input sequence of embeddings $X \in \mathbb{R}^{n \times d}$, the self-attention mechanism computes a weighted sum of these embeddings, allowing each element in the sequence to focus on different parts of the sequence. The self-attention mechanism is defined as the following:

\begin{enumerate}
    \item \textbf{Compute Queries, Keys, and Values:}
    \begin{align*}
        Q &= X W^Q \\
        K &= X W^K \\
        V &= X W^V
    \end{align*}
    where $W^Q$, $W^K$, and $W^V$ are learned weight matrices for queries, keys, and values respectively.

    \item \textbf{Calculate Attention Scores:}
    \begin{equation*}
        S = \frac{Q K^T}{\sqrt{d_k}}
    \end{equation*}
    where $d_k$ is the dimension of the key vectors.

    \item \textbf{Apply Softmax to Scores:}
    \begin{equation*}
        A = \text{softmax}(S)
    \end{equation*}
\end{enumerate}  
\end{definition}
For the transformer model, they do not directly use the Scaled Dot Product Attention. Instead, They split the input sequence of embedding into $h$ parts and do scaled dot product attention at each part in parallel. Then they take a weighted average of it. More specifically, it is defined as the following:
\begin{definition}
 \textbf{Multi Head Attention}   Given an input sequence of embeddings $X$, the multi-head attention mechanism can be described as follows:

\begin{enumerate}
    \item \textbf{Linear Projections:} For each head $i$ from 1 to $n$:
    \begin{align*}
        Q_i &= X W^Q_i \\
        K_i &= X W^K_i \\
        V_i &= X W^V_i
    \end{align*}
    where $W^Q_i$, $W^K_i$, and $W^V_i$ are weight matrices specific to the $i$-th attention head.

    \item \textbf{Compute Scaled Dot-Product Attention for each head:}
    \begin{align*}
        S_i &= \frac{Q_i K_i^T}{\sqrt{d_k}} \\
        A_i &= \text{softmax}(S_i) \\
        H_i &= A_i V_i
    \end{align*}

    \item \textbf{Concatenate and Linearly Transform:}
    \begin{equation*}
        \text{MultiHead}(Q, K, V) = \text{Concat}(H_1, H_2, \dots, H_n) W^O
    \end{equation*}
    where $W^O$ is a learned weight matrix to produce the final output of the multi-head attention.
\end{enumerate}
\end{definition}
Finally, we also need to use the definition of the 
Position-wise Feed-Forward Networks.  
\begin{definition} For a given position $i$ in the sequence, Position-wise Feed-Forward Networks is defined as the following 

\begin{enumerate}
    \item \textbf{First Linear Transformation:}
    \begin{equation*}
        \text{FFN}_{in}(x_i) = x_i W_1 + b_1
    \end{equation*}
    where $W_1$ is a weight matrix and $b_1$ is a bias vector.

    \item \textbf{Activation Function:}
    \begin{equation*}
        \text{FFN}_{\text{ReLU}}(x_i) = \text{ReLU}(\text{FFN}_{in}(x_i))
    \end{equation*}
    where $\text{ReLU}(x) = \max(0, x)$ is the Rectified Linear Unit activation function.

    \item \textbf{Second Linear Transformation:}
    \begin{equation*}
        \text{FFN}_{out}(x_i) = \text{FFN}_{\text{ReLU}}(x_i) W_2 + b_2
    \end{equation*}
    where $W_2$ is another weight matrix and $b_2$ is another bias vector.
\end{enumerate}
 
\end{definition}
Finally, the transformer-based model performs the following process
\begin{enumerate}
    \item \textbf{{Encoder Process}}     The encoder is composed of a stack of 6 layers, with each layer consisting of 2 sub-layers. The first sub-layer is called the multi-head self-attention layer and the second layer is the normal positional-wise fully connected feed-forward network. After going through each layer, it will do normalizations and then produce an output vector of 512 dimensions.
    \item \textbf{Decoder Process}  The decoder is also composed of a stack of 6 layers. The decoder has another multi-head self-attention layer after its first multi-head self-attention layer, followed by the feed-forward network. It also has normalization after each layer.
    \item Finally, it will go through the linear function and softmax function.
\end{enumerate}
\subsubsection{BERT}: As we mentioned before, BERT is a transformer-based model that takes input as bidirectional representations from the unlabeled text by jointly conditioning on both left and right contexts in all layers. And then do the process for Transformer. For more details on how BERT is designed please see~\cite{devlin2018bert}.
\subsection{Classifcation by Feature Vector}

The BERT model will input the text and output a vector called a feature vector. This feature vector is a quantization of the text which contains both semantic and order information. Then after extracting those vectors, we can create a label for each vector by calculating the VBL from \textbf{Definition} \ref{stock_price_tendency} or \textbf{Definition} \ref{stock_price_tendency_relative}. Next we can formulate the classification problem by formulating it as an optimization problem. For this paper, we consider the following 3 optimization problems.
\begin{itemize}
    \item \textbf{Support Vector Machine with regulizar}: The $\ell_2$ and $\ell_1$ regularized support vector machine is formulated as

\begin{equation}
    \min_{w}   \frac{1}{N}\sum_{ i = 1}^N \max(0,1-y_i x_{i}^{T} w_i) + \frac{\mu_1}{2}||w||_{2}{2} + \mu_2 \Vert w \Vert_1
\end{equation}

where $\{(x_i, y_i)\}_{i =1}^{N} $ are given training data with each $y_i \in \{ -1, 1 \}$. Let $w^{*}$  be the solution. Then for a new data point x, it can be classified as $\mbox{sign}(x^{T}w^{*})$ where $\mu_1$ and $\mu_2$ are real values that are non-negative and could be 0.  The support vector machine problem can be generalized to nonlinear classification by using the kernel trick.
\item \textbf{Logistic Regression with regulizar}: The $\ell_2$ regularized logistic regression is formulated as

\begin{equation}
    \min_{w}   \frac{1}{N}\sum_{ i = 1}^N \log(1+\exp(-y_i x_{i}^{T} w)) + \frac{\mu}{2}||w||_{2}{2}
\end{equation}

where $\{(x_i, y_i)\}_{i =1}^{N} $ are given training data with each $y_i \in \{ -1, 1 \}$. Let $w^{*}$  be the solution. Then for a new data point x, it can be classified as $\mbox{sign}(x^{T}w^{*})$. where $\mu$ is a non-negative real value that could be 0.
\item \textbf{Neural Network}: The neural network with $\ell_1$ and $\ell_2$ can be modeled as the following optimization problem: 
\begin{equation}
\min_{\theta} \sum_{i=1}^{N} \ell(f_{\theta}(x_i), y_i) + \mu_1 \lVert \theta \rVert_1 + \frac{\mu_2}{2} \lVert \theta \rVert_2^2
\end{equation}
where $\ell$ is a loss function such as the logarithmic softmax function; $\mu_1$ and $\mu_2$ are real values that are non-negative and could be 0; $f_\theta$ denote a neural network parameterized by $\theta$. The function $f_\theta $ is a composite of several functions of the following form:
\begin{align*}
   \sigma_n(\theta_n \sigma_{n-1}\ \cdots  \sigma_2( \theta_2 \sigma_1(\theta_1 x)))
\end{align*}
And $\sigma_i$ is usually called an activated function. 
\end{itemize}

Now after formulating the classification as an optimization problem, we will also need to use optimization methods to slove. Logistic regression and Support Vector Machine are classical ways to deal with convex optimization problem and their global optimal solutions are tractable; however, for general neural network problems, the optimization problems are nonconvex and only the local optimal solutions are traceable. For our paper, we used the most common method in machine learning which is stochastic gradient descent \cite{amari1993backpropagation}. There are certainly other types of methods can be applied here such as the momentum reduction-based method~\cite{tran2022hybrid,wang2019spiderboost,xu2023momentum}, flow-based methods~\cite{huang2022approximate,wilson2019accelerating,moreau1965proximite} or adaptively based methods \cite{kingma2014adam}. 
\subsection{Topic-Switching Index Calculation}
Having acquired the essential techniques, We will next outline the approach for deriving the Topic-Switching Index. For each earnings call transcript, we extract questions and answers associated with individuals. Unlike previous ltereature, we identify each analyst and pair their questions with the answers. Using BERT(More specifcially we use FineBERT ~\cite{araci2019finbert} which does fine tuning for financial sentiment analysis), we obtain features for each pair. Subsequently, we compute similarity between question and answer vectors from taking their dot product and normalizing by their respective magnitudes. Then we use $1 - \text{similarity score}$ as the Topic-Switching Index for each analyst's Q $\&$ A pair. Finally, we average the Topic-Switching scores and assign the value to this earnings call transcript. 

For classification, each vector is labeled based on the VBL, as determined by either \textbf{Definition} \ref{stock_price_tendency} or \textbf{Definition} \ref{stock_price_tendency_relative}. To train our models — whether it would be a support vector machine, logistic regression, or neural network — we employ stochastic gradient descent. The outcome of this training is a set of weights conducive to classification tasks.

The pseudo-code will be given in \textbf{Algorithm } \ref{algorithm1}
\begin{algorithm}
\caption{Feature and Model Training Process}
\begin{algorithmic}[1]
\Procedure{TrainModel}{EarningsCall}
    \State $TopicSwitchingList \gets \text{empty list}$ \Comment{List to store TopicSwitching}
    
    \For{each earnings call in EarningsCall} \Comment{Iterate through each earnings call}
        \State $TopicSwitching \gets \text{empty list}$
        \For{each question, answer in EarningsCall} 
        \Comment{Process each Q and A}
            \State $feature_{question} \gets \text{FinBERT\_Extract}(question)$ \Comment{Use FinBERT to extract features}
            \State $feature_{answer} \gets \text{FinBERT\_Extract}(answer)$
            
            \State $norm_{question} \gets \lVert feature_{question} \rVert$
            \State $norm_{answer} \gets \lVert feature_{answer} \rVert$
            
            \If{$norm_{question} \neq 0$ and $norm_{answer} \neq 0$} \Comment{Avoid division by zero}
                \State $similarity \gets \frac{feature_{question} \cdot feature_{answer}}{norm_{question} \times norm_{answer}}$
                \State $TopicSwitching.\text{append}( 1- similarity)$ \Comment{Store TopicSwitching}
            \EndIf
        \EndFor
        \State $avgTopicSwitching \gets \text{average}(TopicSwitching)$ \Comment{Compute average TopicSwitchingIndex}
        \State $TopicSwitchingList.\text{append}(avgTopicSwitching)$
    \EndFor

    \If{Calculate feature using \textbf{Definition} \ref{stock_price_tendency} \Comment{Assign label based on definition}}
    \For{each stock in CompantList}
            \State $label \gets \text{VBL according to \textbf{Definition} \ref{stock_price_tendency}}$
    \EndFor
    \EndIf
        \If{Calculate feature using \textbf{Definition} \ref{stock_price_tendency_relative} }
    \For{each stock in CompantList}
            \State $label \gets \text{VBL according to \textbf{Definition} \ref{stock_price_tendency_relative}}$
    \EndFor
    \EndIf
    \State Initialize a model (SVM, Logistic Regression, or Neural Network)
    \State Use stochastic gradient descent to train the model with features and labels
    \State $weights \gets$ trained model's weights
    \State \Return $weights$
\EndProcedure
\end{algorithmic}
\label{algorithm1}
\end{algorithm}

\section{Data}
\subsection{Data Overview}
We fetch the earnings call data of the $S\& P$ 500 companies from the website https://seekingalpha.com and get a total amount of 24,573 transcripts. Our transcript dataset ranges from 2010 January to 2022 December. After excluding data that have missing values such as not providing the NYSE symbols, the text of which does not have the same format with others and the content cannot be separate, or other miscellaneous problems that hard to fix, 13,044 ones are left and can successfully calculate the Topic-Switching Index. We also get the corresponding stock prices before and after the earnings call day from yahoo finance and match them into the whole dataset.
\subsection{Descriptive Analysis}
In this subsection, our objective is to delve deeper into the intricacies of the data associated with the Topic-Switching Feature by showing visualizations. 

\begin{figure}
    \centering
    \includegraphics[width=\linewidth]{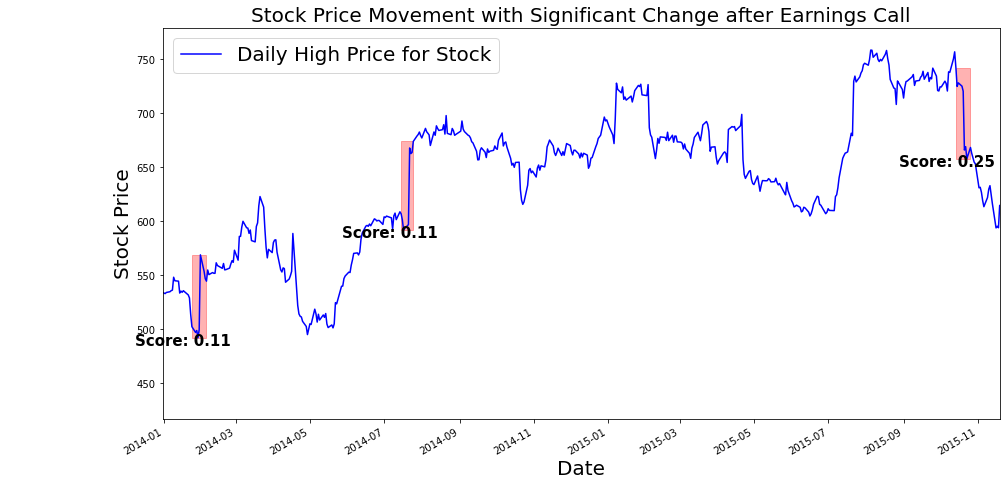}
    \caption{Stock Price movement of Chipotle Mexican Grill}
    \label{fig:Stock_Price_Sample1}
\end{figure}

To have a more clear and practical understanding of our index, We show 2 examples of discussion text in the \textbf{Appendix A1}, one with a low Topic Switching Index and one with a high score. From the low score sample, we can see that the manager answered the analyst to the point with confidence and all the details put were for the arguements. While in the other one, the manager provided context and touched upon various related topics but did not give a straightforward answer about what the "new normal" might be regarding investment in ex-fuel gross margin which the analyst was interested in. After checking samples with scores, we preliminarily can see that our methodology does extract some information based on manager's attempt to evade from answering questions directly and assign it into Topic-Switching Index. Moreover, we also show the scores on the same plot as in the previous part for the earnings calls that witness a huge jump of Chipotle Mexican Grill stock price. Notice that the average score for the whole sample is $0.24$. The two upwards jump are with Topic-Switching Index of 0.11 and the downward one has a score of 0.25. It is clear from figure \ref{fig:Stock_Price_Sample1} that the stock price moves with an opposite direction with Topic-Switching Index in this case.

\begin{figure}
    \centering
    \begin{subfigure}{\textwidth}
        \centering
        \includegraphics[width=\linewidth]{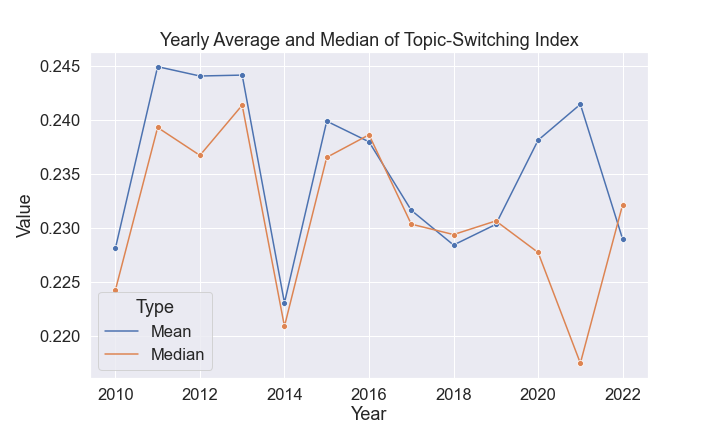}
        \caption{Yearly Trend for Topic-Switching Index}
    \end{subfigure}%
    \hfill
    \begin{subfigure}{\textwidth}
        \centering
        \includegraphics[width=\linewidth]{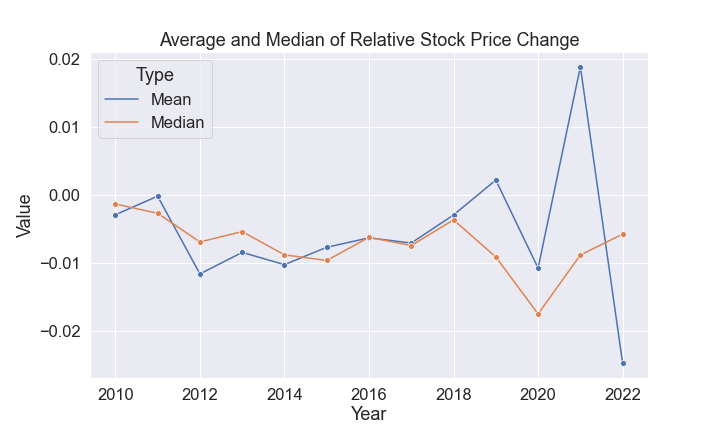}
        \caption{Yearly Trend for Relative Stock Price Change}
    \end{subfigure}
    \caption{Yearly Trend for Topic-Switching Index and Relative Stock Price Change}
    \label{fig:Yearly_Trend}
\end{figure}

In figure \ref{fig:Yearly_Trend}, we show the yearly trends for Topic-Switching Index and stock price change of the whole sample. For each plot, we include both median and mean values. For Topic-Switching Index, on average the mean is higher than the median, indicating the data are skewed to the right. The abnomal differences appear on year 2020 and 2021 might come from the Covid situation. Still, the difference is only around 0.025. There is no clear yearly trend for the index. For relative daily stock price change, it looks like the mean and median do not have a real difference except during Covid period. In order to be consistent, we will use median in our later research.

To structure our exploration, we categorize the entire dataset into 11 distinct segments: Consumer Discretionary, Health Care, Information Technology, Consumer Staples, Industrials, Communication Services, Financials, Materials, Energy, Real Estate, and Utilities based on the Global Industry Classification Standard (GICS). This categorization can allow for a more organized and insightful analysis. Figure \ref{fig:Box_Plot} below shows the box plots of Topic-Switching Index and relative stock price change for each category. The color is based on the rank of median in both plots, the higher the median of the industy, the darker its box appear. The boundaries of the whiskers are based on the 1.5 interquartile range. For Topic-Switching Index, we can see that while Materials having the highest median and Communication Services ranking the lowest, overall the medians are every close to each other, being around 0.25. The whiskers are similar also. For the relative daily stock price change, the medians are all close to 0 as expected. Although there is no strong evidence that the rank of industry's index median is related to that of its stock price daily change, it is understandable as the difference among industries are insignificant.

\begin{figure}
    \centering
    \begin{subfigure}{\textwidth}
        \centering
        \includegraphics[width=\linewidth]{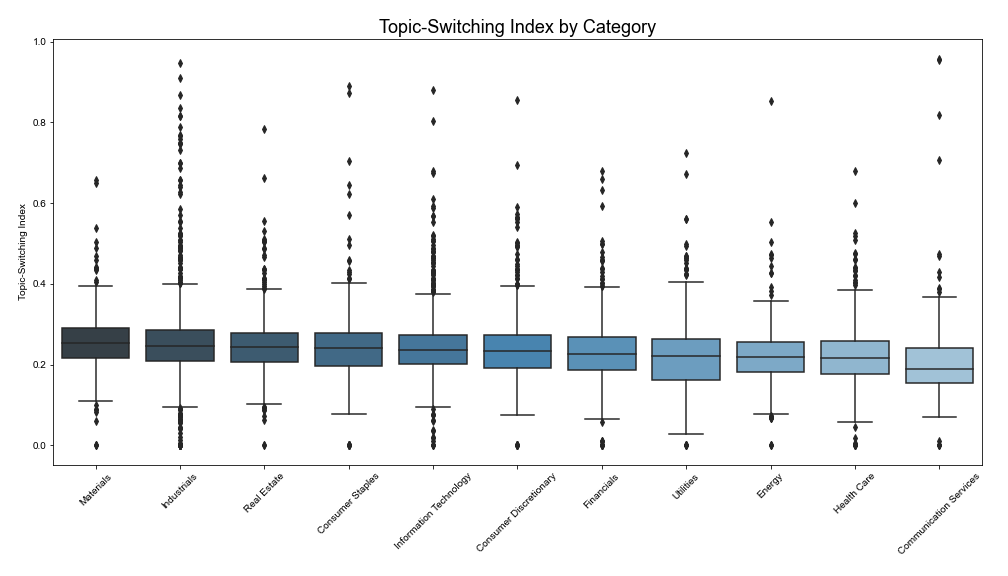}
        \caption{Box Plot for Topic-Switching Index for Each Category}
    \end{subfigure}
    \begin{subfigure}{\textwidth}
        \centering
        \includegraphics[width=\linewidth]{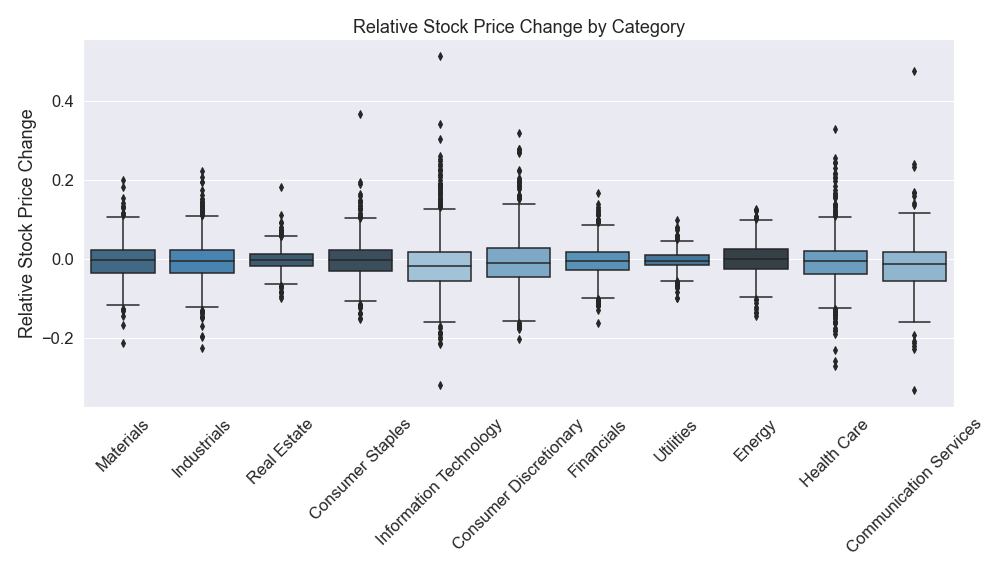}
        \caption{Box Plot for Stock Price Change for Each Category}
    \end{subfigure}
    \caption{Box plot for Topic-Switching Index and relative stock price change.}
    \label{fig:Box_Plot}
\end{figure}
In addition, we show the statistics of Topic-Switching Index of all categories in \textbf{Table} \ref{basic_stat}, including the maximum, minimum, mean, standard deviation, and overall data count. As we have observed in the box plot, the distribution of each segment does not distinguish with each other significantly with an average close to 0.25. 
\begin{table}[H] 
\centering
\caption{Statistics for Topic-Switching Index of All Catgories}
\begin{tabular}{lcccccc} 
\toprule
Category & Mean & Std Dev & Total Data Points & Minimum & Maximum \\
\midrule
Consumer Discretionary & 0.24& 0.07 & 1387 & 0 & 0.85 \\
Health Care & 0.22 & 0.07 & 1592 &0 & 0.68 \\
Information Technology &0.24 & 0.07 & 1647 & 0 & 0.88 \\
Consumer Staples & 0.24 & 0.07 & 941 & 0 & 0.89 \\
Industrials & 0.25 & 0.09 & 1710 &0 & 0.95 \\
Communication Services & 0.21 & 0.09 & 407 & 0& 0.96 \\
Financials & 0.23 & 0.07 & 1764 &0 & 0.68 \\
Materials & 0.25 & 0.07 & 664 & 0 & 0.66 \\
Energy & 0.22 & 0.07 & 558 &0 & 0.85\\
Real Estate & 0.25 & 0.07 & 750 & 0& 0.78 \\
Utilities & 0.22 & 0.09 & 751 & 0 & 0.72\\
\bottomrule
\end{tabular}
\label{basic_stat}
\end{table}

\section{Experiment}
\subsection{Linear Regression}
In the following analysis, we will first employ basic linear regression to preliminarily represent the relationship between the Topic-Switching Index and the relative daily change in stock price. 

First, let us introduce some notation to facilitate our discussion. Consider \( \xi_{d}^{c} \) to represent the Topic-Switching Feature of a given company \( c \) on day \( d \). Concurrently, the relative change in stock price can be captured by the expression \( \frac{S_{d+1}^c - S_{d-1}^{c}}{S_{d-1}} \). Here, \( S_d^c \) stands for the stock price of the company \( c \) at time \( d \). 

With the defined variables, our next step is to apply linear regression to see whether there exists a negative correlation between the Topic-Switching Index and the relative stock price change. The results are shown in \textbf{Table} \ref{basic_stat_linear}.

From the table, we can see that 9 out of 11 industries have nagative coeeficients and the whole sample itself also shows a negative correlation. The sector Materials has a positive coefficient both statistically and economically insignificant. Moreover, this sector has a relative small number of total data points as presented in \textbf{Table} \ref{basic_stat}. The coeffient for Industrials is also positive, which can be explained its unique data distribution. From the first box plot in Figure \ref{fig:Box_Plot}, we can see that the Topic-Switching scores for Industrials are very dispersed, with many of them being out of the 1.5 Interquartile range and way above. By doing linear regression based on the whole sample without segmentation, we have an estimated parameter of -2.07$\%$ with a t-value of -3.426. With the above analysis, we are confident to say that there exists a negative correlation between our Topic-Switching Index and the stock price movement, meaning that the market does capture manager's suspicious talking strategy and reflect that in the stock price.

Moving forward to the classifier models section, we plan to explore this hypothesis in depth. Specifically, we will focus on the predictability on the stock price movement using the Topic-Switching Feature as a key metric.
\begin{table}[H] 
\centering
\caption{Linear Regression Statistic for All Categories} 
\begin{tabular}{|l|c|c|c|}
\toprule
                Sector &  Coefficient &  Std. Error &  t-value \\
\midrule
Consumer Discretionary &      -0.0665 &      0.0234 &  -2.8363 \\
           Health Care &      -0.0237 &      0.0203 &  -1.1654 \\
Information Technology &      -0.0547 &      0.0229 &  -2.3867 \\
      Consumer Staples &      -0.0150 &      0.0201 &  -0.7469 \\
           Industrials &       0.0248 &      0.0125 &   1.9836 \\
Communication Services &      -0.0541 &      0.0366 &  -1.4780 \\
            Financials &      -0.0236 &      0.0132 &  -1.7865 \\
             Materials &       0.0004 &      0.0276 &   0.0128 \\
                Energy &      -0.0003 &      0.0251 &  -0.0106 \\
           Real Estate &      -0.0313 &      0.0140 &  -2.2386 \\
             Utilities &      -0.0040 &      0.0097 &  -0.4070 \\
               Overall &      -0.0207 &      0.0060 &  -3.4264 \\
\bottomrule
\end{tabular}

\label{basic_stat_linear}
\end{table}

\subsection{Classifier Models}
In this section, we use the 3 different classifier models as we mentioned before (SVM, Logistic Regression, and Neural Network) to illustrate the predictive power of the Topic-Switching Index. We will compare our index to the benchmark feature which is directly using the transformer model to extract the text feature from the earnings call. We also compare the results with the comprehensive feature which adds Topic-Switching Index to the text feature. We separate our data into two parts, with training data being all the data before 2016 and the rest being the testing data. 

The results of which the label of the classification is calculated by \textbf{Definition }\ref{stock_price_tendency} are shown in Table 3. From this table, it is clear that using Topic-Switching Index alone has the best out of sample accuracy for all models. Three models themselves do not really distinguish with each other by performance with Logistic Regression has a  small advantage over others.

\begin{table}
\centering
\caption{The testing accuracy by using \textbf{Definition }\ref{stock_price_tendency} }
\begin{tabular}{|l|c|c|c|c|}
\toprule
 & Support Vector Machine & Logistic Regression & Neural Network \\
\hline
\hline
Benchmark Feature & 0.540 & 0.542 & 0.521 \\
\hline
\hline
Benchmark with Topic-Switching Index & 0.539 & 0.544 & 0.543 \\
\hline
\hline
 Topic-Switching Index & \textbf{0.570} & \textbf{0.570} & \textbf{0.570} \\
\hline 
\bottomrule
\end{tabular}
\end{table}

\begin{table}
\centering
\caption{The testing accuracy by using \textbf{Definition }\ref{stock_price_tendency_relative}  and $\tau = -0.01$}
\begin{tabular}{|l|c|c|c|c|}
\toprule
 & Support Vector Machine & Logistic Regression & Neural Network \\
\hline
\hline
Benchmark Feature & 0.525 & 0.519 & 0.513 \\
\hline
\hline
Benchmark with Topic-Switching Index & 0.526 & 0.521 & 0.508 \\
\hline
\hline
 Topic-Switching Index & \textbf{0.548} & \textbf{0.548} & \textbf{0.548} \\
\hline 
\bottomrule
\end{tabular}
\end{table}

\begin{table} 
\centering
\caption{The testing accuracy by using \textbf{Definition }\ref{stock_price_tendency_relative}  and $\tau = -0.02$}
\begin{tabular}{|l|c|c|c|c|}
\toprule
 & Support Vector Machine & Logistic Regression & Neural Network \\
\hline
\hline
Benchmark Feature & 0.627 & 0.611 & 0.564 \\
\hline
\hline
Benchmark with Topic-Switching Index & 0.627 & 0.614 & 0.597 \\
\hline
\hline
 Topic-Switching Index & \textbf{0.642} & \textbf{0.642} & \textbf{0.642} \\
\hline 
\bottomrule
\end{tabular}
\end{table}
\begin{table}
\centering
\caption{The testing accuracy by using \textbf{Definition }\ref{stock_price_tendency_relative}  and $\tau = -0.05$}
\begin{tabular}{|l|c|c|c|c|}
\toprule
 & Support Vector Machine & Logistic Regression & Neural Network \\
\hline
\hline
Benchmark Feature & \textbf{0.839} & 0.828 & 0.799 \\
\hline
\hline
Benchmark with Topic-Switching Index & \textbf{0.839} & 0.828 & 0.776 \\
\hline
\hline
 Topic-Switching Index & \textbf{0.839} & \textbf{0.839} & \textbf{0.839} \\
\hline 
\bottomrule
\end{tabular}
\end{table}
\begin{table}
\centering
\caption{The testing accuracy by using \textbf{Definition }\ref{stock_price_tendency_relative}  and $\tau = 0.01$}
\begin{tabular}{|l|c|c|c|c|}
\toprule
 & Support Vector Machine & Logistic Regression & Neural Network \\
\hline
\hline
Benchmark Feature & 0.6627 & 0.643 & 0.595 \\
\hline
\hline
Benchmark with Topic-Switching Index & 0.6627 & 0.642 & 0.595 \\
\hline
\hline
 Topic-Switching Index & \textbf{0.6629} & \textbf{0.6629} & \textbf{0.6629} \\
\hline 
\bottomrule
\end{tabular}
\end{table}

\begin{table}
\centering
\caption{The testing accuracy by using \textbf{Definition }\ref{stock_price_tendency_relative}  and $\tau = 0.02$}
\begin{tabular}{|l|c|c|c|c|}
\toprule
 & Support Vector Machine & Logistic Regression & Neural Network \\
\hline
\hline
Topic-Switching Index & \textbf{0.740} & 0.731 & 0.701 \\
\hline
\hline
Benchmark with Topic-Switching Index & \textbf{0.740} & 0.731 & 0.701 \\
\hline
\hline
 Topic-Switching Index & \textbf{0.740} & \textbf{0.740} & \textbf{0.740} \\
\hline 
\bottomrule
\end{tabular}
\end{table}

\begin{table}
\centering
\caption{The testing accuracy by using \textbf{Definition }\ref{stock_price_tendency_relative}  and $\tau = 0.05$}
\begin{tabular}{|l|c|c|c|c|}
\toprule
 & Support Vector Machine & Logistic Regression & Neural Network \\
\hline
\hline
Benchmark Feature & \textbf{0.888} & 0.886 & 0.878 \\
\hline
\hline
Benchmark with Topic-Switching Index & \textbf{0.888} & 0.888 & 0.879 \\
\hline
\hline
 Topic-Switching Index & \textbf{0.888} & \textbf{0.888} & \textbf{0.888} \\
\hline 
\bottomrule
\end{tabular}
\end{table}

Table 4-9 show the results based on \textbf{Definition} \ref{stock_price_tendency_relative}. Table 4-6 predict the relative stock price goes down more than 1$\%$, 2$\%$ and 5$\%$ while Table 7-9 predict the price jumps up by more than 1$\%$, 2$\%$ and 5$\%$. According to the results, we notice that the Topic-Switching Index outperforms the Benchmark Feature and Benchmark with Topic-Switching Index in all cases. Additionally, there is no consistent major difference between the Topic-Switching with Benchmark Features and the Benchmark Feature.

The best performance of using only Topic-Switching Index indicates that our index has more predictive power than feeding the whole texts to FineBERT model. 
In addition, because the Topic-Switching Index is only a one-dimensional vector, it has less overfitting problem, which explains why the index alone outperforms the index with Benchmark Features.
We also notice that the Topic-Switching Index is more robust over different models, which is observed from the fact that it has very similar accuracy with respect to different models, and is also resulting from that the feature has only one dimension.
The accuracy of all methods increases when using \textbf{Definition} \ref{stock_price_tendency_relative} and smaller(larger) $ \tau $. 
This is indicates that the percentage of stocks that have such a relative increase that will converge to $100 \%$ (0 $\%$) as $\tau \to +\infty$($-\infty$), hence it is natural to guess that when the stock price increase(decrease) relatively, all models will capture this information with all different features. This can also explain why the accuracy of predicting higher jumps is better than downward drops as there are fewer big upward jumps in the sample and is easier to predict.

\section{Conclusion and Future Work}
In this paper, we first introduce a new feature, topic-switching, to consider in analyzing earnings call transcripts. We then provide a novel way to quantify this feature based on a transformer model called FinBERT. The logic is that when the manager is trying to evade from answering the question from the analyst directly, the similarity between both parties' contents becomes lower, and we try to quantify the part where the question and answer do not overlap. 
By looking into the transcripts and Topic-Switching Index both visually and manually, we find that our calculated index does capture this information. Next, we show that the stock price movement is nagatively correlated to the Topic-Switching Index, confirming that the investors notice the evasive talking strategy adopted by the manager and show their suspicion in the investment choice. Moreover, we demonstrate that this feature has predictive power and is better than using FinBERT directly, which is our benchmark. 

There is still much future work left to be done. It will be more helpful if our hyposethis can be tested on a larger dataset including companies not only listed on S$\&$P 500. In addtion, we will expand our testing time range to ckeck whether this Topic-Switching Index have effects on the stock price for a longer period such as a week. The next step of our research is to examine whether the market is over-reacting to manager's evasion in firm disclosure by looking into the predictablility of our index to the company's future performance.
\newpage
\appendix
\section{Appendix}
\subsection{Sample Tanscript}
The following is a sample of transcript with Low Topic Switching Score:

"\textit{Shannon Cross: 
Tim, can you talk a bit about what you are seeing in China with 70$\%$ year over year growth in the Greater China revenue, and clearly very strong iPhone. If you can talk a bit about what consumers are saying, what the carriers are saying, in terms of demand and opportunity. Just any color, because clearly itis quite strong.
}

\textit{
Tim Cook:
Yeah, it was an incredible quarter. We were up 71$\%$ year over year. We set a record in China for revenues. We did that now in a quarter that included Chinese New Year, and so we have the help of a strong holiday season, much like the U.S. has a strong season in December. China is obviously in the March quarter.
iPhone led the way. It was up over 70$\%$ year on year. And the current estimates from Kantar are that that would mean that we would gain more than 9 points of share on a year over year basis. And so by everything I can see, we did extremely well.
The Mac also had an unbelievable quarter in China, and I am particularly very happy with this, that Mac unit sales were up 31$\%$. And like the rest of the world, or most of the rest of the world, IDC is projecting that PC sales in China contracted by 5$\%$ last quarter. And so once again bucking the tide.
Also, in China, consistent with the company but at a much different rate, the App Store had a record quarter and grew over 100$\%$ year over year. And so you can see the iPhone, the Mac, and the App Store adding, and with the iPad in PRC, not in Greater China, but in the PRC, iPad had its best quarter ever, higher than all the others, and also grew in a market that contracted for the overall market.
And so really and truly, itis sort of everything you look at in China was extremely good. We have been working significantly on expanding our ecosystem there, and so we added Union Pay as a payment option for customers. We increased the iPhone point of sales to over 40,000 during the quarter. Thatis up about 9$\%$ year on year. And more importantly than the total number, we are in many more cities than we were before.
We worked significantly on our online store. Our online store revenue was up over three times year over year. As you probably heard us say before, we have opened several stores in China recently. We are now at 21 in Greater China and we are on track still to achieve 40 stores by the middle of next year. The online store will also be expanding from around 319 cities to where they can hit two day delivery to 365 cities. So adding about 50 new cities by the end of this quarter.
And so the net is we are investing a lot across the board in our infrastructure, in our products, on partnering with different companies. The Chinese developers are coming on in significant numbers. We have now made payments to developers in Greater China of almost 5 billion over half of which was in the last 12 months.
And so you can see this enormous momentum building in the developer community there as well. And so lots of positive things, and you know, as you probably heard me say before, I have never seen as many people coming into the middle class as they are in China. And that is where the bulk of our sales are going. And so we are really proud of the results there and continue to invest in the country.}"
\newline

The second example is a sample of transcript with high Topic Switching Score:

"\textit{
Judah Frommer
Okay, that makes sense. And then touching on the gross margin performance, I mean you have lapped Harris Teeter fully now and we still don't see a lot of investment in margin. And maybe Mike referenced it in that you're doing good work with 84.51, but is this kind of a new normal, like a less than 10 basis point investment in ex-fuel gross margin?}

\textit{
J. Michael Schlotman
I'll hesitate to give what a normal is for an investment because we do continue to make investments in price in a variety of areas of our store. I'd say it's a reflection of prudence on how and when we make investments. Additionally, the mix, when you look at the strength of brands like Simple Truth, when you look at the strength of natural and organics, all of which have a little bit higher gross margin, that certainly plays into the overall mix of business. But if you were to look at gross margins by department, I have a little bit different view of exactly how and when we invest. But we will continue to invest not only in price but all four keys of our Customer 1st strategy.
}"
\bibliographystyle{unsrt}
\bibliography{refs}

\begin{thebibliography}{10}

\bibitem{matsumoto2011makes}
Dawn Matsumoto, Maarten Pronk, and Erik Roelofsen.
\newblock What makes conference calls useful? the information content of
  managers' presentations and analysts' discussion sessions.
\newblock {\em The Accounting Review}, 86(4):1383--1414, 2011.

\bibitem{chen2018manager}
Jason~V Chen, Venky Nagar, and Jordan Schoenfeld.
\newblock Manager-analyst conversations in earnings conference calls.
\newblock {\em Review of Accounting Studies}, 23:1315--1354, 2018.

\bibitem{kimbrough2005effect}
Michael~D Kimbrough.
\newblock The effect of conference calls on analyst and market underreaction to
  earnings announcements.
\newblock {\em The Accounting Review}, 80(1):189--219, 2005.

\bibitem{garcia2023colour}
Diego Garcia, Xiaowen Hu, and Maximilian Rohrer.
\newblock The colour of finance words.
\newblock {\em Journal of Financial Economics}, 147(3):525--549, 2023.

\bibitem{ke2019predicting}
Zheng~Tracy Ke, Bryan~T Kelly, and Dacheng Xiu.
\newblock Predicting returns with text data.
\newblock Technical report, National Bureau of Economic Research, 2019.

\bibitem{hollander2010does}
Stephan Hollander, Maarten Pronk, and Erik Roelofsen.
\newblock Does silence speak? an empirical analysis of disclosure choices
  during conference calls.
\newblock {\em Journal of Accounting Research}, 48(3):531--563, 2010.

\bibitem{youngs2020linguistic}
Johnathan Youngs.
\newblock Linguistic complexity and the post-earnings announcement drift.
\newblock {\em DU Undergraduate Research Journal Archive}, 1(2):7, 2020.

\bibitem{chi2022does}
Danlin Chi, Hasibul Chowdhury, Nicolas Eugster, and Jiayi Zheng.
\newblock Does linguistic complexity of annual reports affect corporate leasing
  decision?
\newblock {\em Available at SSRN 4214982}, 2022.

\bibitem{bushee2021analysts}
Brian~J Bushee and Ying~Julie Huang.
\newblock Do analysts and investors efficiently respond to managerial
  linguistic complexity on conference calls?
\newblock {\em Available at SSRN 3405086}, 2021.

\bibitem{li2008annual}
Feng Li.
\newblock Annual report readability, current earnings, and earnings
  persistence.
\newblock {\em Journal of Accounting and economics}, 45(2-3):221--247, 2008.

\bibitem{bloomfield2008discussion}
Robert Bloomfield.
\newblock Discussion of “annual report readability, current earnings, and
  earnings persistence”.
\newblock {\em Journal of Accounting and Economics}, 45(2-3):248--252, 2008.

\bibitem{guay2016guiding}
Wayne Guay, Delphine Samuels, and Daniel Taylor.
\newblock Guiding through the fog: Financial statement complexity and voluntary
  disclosure.
\newblock {\em Journal of Accounting and Economics}, 62(2-3):234--269, 2016.

\bibitem{dyer2017evolution}
Travis Dyer, Mark Lang, and Lorien Stice-Lawrence.
\newblock The evolution of 10-k textual disclosure: Evidence from latent
  dirichlet allocation.
\newblock {\em Journal of Accounting and Economics}, 64(2-3):221--245, 2017.

\bibitem{bushee2018linguistic}
Brian~J Bushee, Ian~D Gow, and Daniel~J Taylor.
\newblock Linguistic complexity in firm disclosures: Obfuscation or
  information?
\newblock {\em Journal of Accounting Research}, 56(1):85--121, 2018.

\bibitem{gow2021non}
Ian~D Gow, David~F Larcker, and Anastasia~A Zakolyukina.
\newblock Non-answers during conference calls.
\newblock {\em Journal of Accounting Research}, 59(4):1349--1384, 2021.

\bibitem{lee2016can}
Joshua Lee.
\newblock Can investors detect managers' lack of spontaneity? adherence to
  predetermined scripts during earnings conference calls.
\newblock {\em The Accounting Review}, 91(1):229--250, 2016.

\bibitem{davis2015effect}
Angela~K Davis, Weili Ge, Dawn Matsumoto, and Jenny~Li Zhang.
\newblock The effect of manager-specific optimism on the tone of earnings
  conference calls.
\newblock {\em Review of Accounting Studies}, 20:639--673, 2015.

\bibitem{devlin2018bert}
Jacob Devlin, Ming-Wei Chang, Kenton Lee, and Kristina Toutanova.
\newblock Bert: Pre-training of deep bidirectional transformers for language
  understanding.
\newblock {\em arXiv preprint arXiv:1810.04805}, 2018.

\bibitem{liu2021finbert}
Zhuang Liu, Degen Huang, Kaiyu Huang, Zhuang Li, and Jun Zhao.
\newblock Finbert: A pre-trained financial language representation model for
  financial text mining.
\newblock In {\em Proceedings of the twenty-ninth international conference on
  international joint conferences on artificial intelligence}, pages
  4513--4519, 2021.

\bibitem{araci2019finbert}
Dogu Araci.
\newblock Finbert: Financial sentiment analysis with pre-trained language
  models.
\newblock {\em arXiv preprint arXiv:1908.10063}, 2019.

\bibitem{medya2022exploratory}
Sourav Medya, Mohammad Rasoolinejad, Yang Yang, and Brian Uzzi.
\newblock An exploratory study of stock price movements from earnings calls.
\newblock In {\em Companion Proceedings of the Web Conference 2022}, pages
  20--31, 2022.

\bibitem{vaswani2017attention}
Ashish Vaswani, Noam Shazeer, Niki Parmar, Jakob Uszkoreit, Llion Jones,
  Aidan~N Gomez, {\L}ukasz Kaiser, and Illia Polosukhin.
\newblock Attention is all you need.
\newblock {\em Advances in neural information processing systems}, 30, 2017.

\bibitem{cho2014learning}
Kyunghyun Cho, Bart Van~Merri{\"e}nboer, Caglar Gulcehre, Dzmitry Bahdanau,
  Fethi Bougares, Holger Schwenk, and Yoshua Bengio.
\newblock Learning phrase representations using rnn encoder-decoder for
  statistical machine translation.
\newblock {\em arXiv preprint arXiv:1406.1078}, 2014.

\bibitem{bahdanau2014neural}
Dzmitry Bahdanau, Kyunghyun Cho, and Yoshua Bengio.
\newblock Neural machine translation by jointly learning to align and
  translate.
\newblock {\em arXiv preprint arXiv:1409.0473}, 2014.

\bibitem{sutskever2014sequence}
Ilya Sutskever, Oriol Vinyals, and Quoc~V Le.
\newblock Sequence to sequence learning with neural networks.
\newblock {\em Advances in neural information processing systems}, 27, 2014.

\bibitem{amari1993backpropagation}
Shun-ichi Amari.
\newblock Backpropagation and stochastic gradient descent method.
\newblock {\em Neurocomputing}, 5(4-5):185--196, 1993.

\bibitem{tran2022hybrid}
Quoc Tran-Dinh, Nhan~H Pham, Dzung~T Phan, and Lam~M Nguyen.
\newblock A hybrid stochastic optimization framework for composite nonconvex
  optimization.
\newblock {\em Mathematical Programming}, 191(2):1005--1071, 2022.

\bibitem{wang2019spiderboost}
Zhe Wang, Kaiyi Ji, Yi~Zhou, Yingbin Liang, and Vahid Tarokh.
\newblock Spiderboost and momentum: Faster variance reduction algorithms.
\newblock {\em Advances in Neural Information Processing Systems}, 32, 2019.

\bibitem{xu2023momentum}
Yangyang Xu and Yibo Xu.
\newblock Momentum-based variance-reduced proximal stochastic gradient method
  for composite nonconvex stochastic optimization.
\newblock {\em Journal of Optimization Theory and Applications},
  196(1):266--297, 2023.

\bibitem{huang2022approximate}
Yushen Huang, Taehoon Lee, and Yifan Sun.
\newblock Approximate backwards differentiation of gradient flow.
\newblock {\em arXiv preprint arXiv:2211.04653}, 2022.

\bibitem{wilson2019accelerating}
Ashia~C Wilson, Lester Mackey, and Andre Wibisono.
\newblock Accelerating rescaled gradient descent: Fast optimization of smooth
  functions.
\newblock {\em Advances in Neural Information Processing Systems}, 32, 2019.

\bibitem{moreau1965proximite}
Jean-Jacques Moreau.
\newblock Proximit{\'e} et dualit{\'e} dans un espace hilbertien.
\newblock {\em Bulletin de la Soci{\'e}t{\'e} math{\'e}matique de France},
  93:273--299, 1965.

\bibitem{kingma2014adam}
Diederik~P Kingma and Jimmy Ba.
\newblock Adam: A method for stochastic optimization.
\newblock {\em arXiv preprint arXiv:1412.6980}, 2014.

\end{thebibliography}
\end{document}